\documentclass[twoside]{article}
\usepackage{fleqn,espcrc2}
\usepackage{graphicx}

\newcommand{\AmS}{{\protect\the\textfont2
  A\kern-.1667em\lower.5ex\hbox{M}\kern-.125emS}}
\title{Sigma pole position and errors of a once and twice subtracted dispersive
analysis of pi-pi scattering data}
\author{R. Kami\'nski\address{Department of Theoretical Physics, H. Niewodnicza\'nski Institute
 of Nuclear Physics, Polish Academy of Sciences, 31-342 
 Krak\'ow, Poland}%
        ,
        R. Garcia-Martin\address{Departamento de F\'{\i}sica Te\'orica,~II
 (M\'etodos Matem\'aticos),
Facultad de Ciencias F\'{\i}sicas,
Universidad Complutense de Madrid,
E-28040, Madrid, Spain},
P. Grynkiewicz\address{ul. Pauli\'nska, 6/18 Krak\'ow 31-065}%
,
J. R. Pel\'aez$^b$%
}

\begin{document}
\begin{abstract}
We show how the new precise data on 
kaon decays together with forward dispersion
relations, sum rules and once- and twice-subtracted
 Roy's equations allow for a precise
determination of the sigma meson pole position. 
We present a comparison and a study of 
the different sources of uncertainties when using either 
once- or twice-subtracted Roy's equations to analyze the data.
Finally we present a preliminary determination of the $\sigma$ pole
from the constrained dispersive data analysis.
\end{abstract}

\maketitle
\section{Introduction}
Roy's equations (RE), based on twice-subtracted dispersion relations and
crossing symmetry conditions for $\pi\pi\to\pi\pi$ amplitudes
were obtained in 1971~\cite{Roy71}.
In recent years, these equations have been used
either to obtain predictions for low energy 
$\pi\pi$ scattering, sometimes using  Chiral Perturbation Theory (ChPT)
\cite{A4,CGLNPB01},
or to test ChPT ~\cite{DescotesGenon:2001tn,KJYII,KJYI}, as well as to
solve old data ambiguities \cite{Kaminski:2002pe}.
Roy eqs. are relevant for the sigma pole,
whose position has also been predicted 
very precisely with the help of ChPT~\cite{Caprini}.

Our group~\cite{KJYII,KJYI} has also used 
Roy eqs. with Forward Dispersion Relations (FDR)
to obtain a precise determination of $\pi\pi$ scattering amplitudes from data
consistent with analyticity, unitarity and crossing. On purpose, we have not
included ChPT constraints, so that we can use our results 
as tests of the ChPT predictions. Unfortunately, 
the large experimental error of the scattering length $a^2_0$ 
of the isospin 2 scalar partial wave, becomes 
a very
large error for the sigma pole determination using RE. 
For this reason, a new set of once-subtracted RE, 
called GKPY eqs. for brevity,
have been derived.
Both the RE and GKPY equations provide analytic extensions 
for the calculation of poles in the complex plane.
Here we 
present preliminary results for the $\sigma$ pole in the S0 wave
obtained from GKPY eqs. 
and analyse the 
different sources of errors comparing their size with those from RE.

\section{Roy versus GKPY Equations}
In practice Roy's equations relate 
real parts (output) of the $S0$, $P$ and $S2$ partial wave amplitudes $f_{\ell}^{I}(s)$ below some energy scale
$s_{max}$,
 with the imaginary parts of all other partial waves and high energy parametrizations (input).
Both GKPY eqs. and RE can be split into three terms:
$$Re f_{\ell}^{I}(s) = ST_{\ell}^{I}(s) + KT_{\ell}^{I}(s) + DT_{\ell}^{I}(s).$$

The ``subtraction terms'' $ST(s)$ are polynomials
whose coefficients are linear combinations of the S0 and S2
scattering lengths. For standard, twice subtracted, Roy's eqs., 
$ST$ are  first degree polynomials,
thus growing quadratically with energy, whereas for GKPY they are 
just constants. Hence, the uncertainty due to the poor
experimental knowledge of the isospin 2 scalar scattering length, $a^2_0$
becomes a very large source of error at high energies for standard
Roy's eqs.

The ``kernel terms'' $KT(s)$ parametrize the contribution of the
$S0$, $P$ and $S2$ waves ($\ell=0,1$)
up to a given energy $s_{max}$,
whereas the ``driving terms'' $DT(s)$ refer to all
waves above $s_{max}$ and also to $\ell>1$ partial waves from threshold.
In our case $\sqrt{s_{max}} = 1420$~MeV and for higher energies we use
Regge parametrizations.
Both the kernel 
and driving terms are integrals of the kind:
\begin{equation}
  \sum_{I'=0}^{2}\sum_\ell
  \int ds'\,
  K_{\ell \ell^\prime}^{I I^\prime}(s,s') \mbox{Im }f_{\ell'}^{I^\prime}(s').
\end{equation}

The integration kernels
$K_{\ell \ell^\prime}^{I I^\prime}(s,s')$ are different 
for RE and GKPY eqs. The important
fact being that, for high $s'$, they 
decrease as $\sim 1/s'^3$ for RE, but as $\sim 1/s'^2$
for GKPY eqs. Hence GKPY eqs. are more sensitive to the high energy input.

\begin{figure}[ht!]
\vspace{9pt}
\includegraphics*[width=6.0cm]{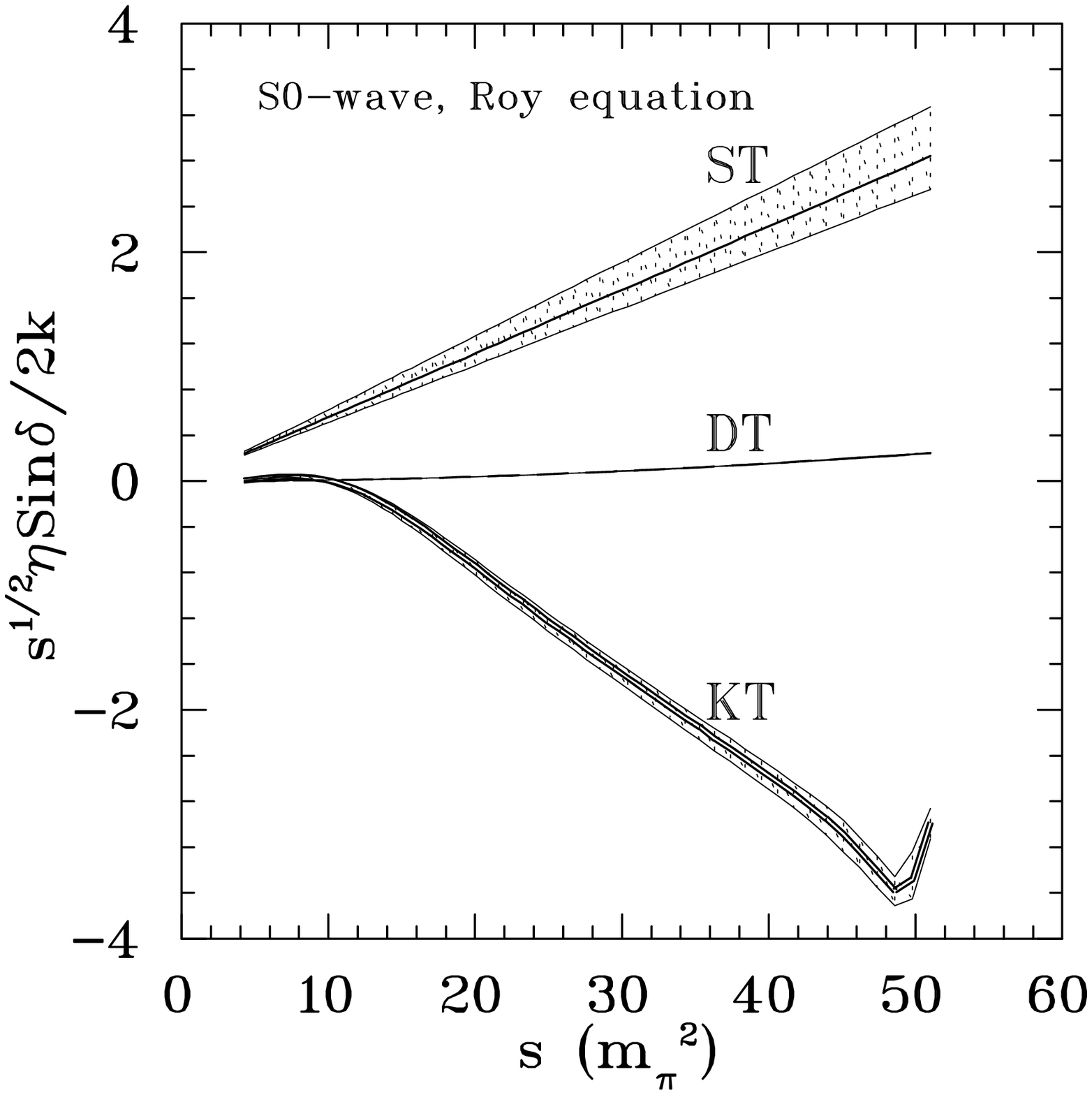}
\includegraphics*[width=6.0cm]{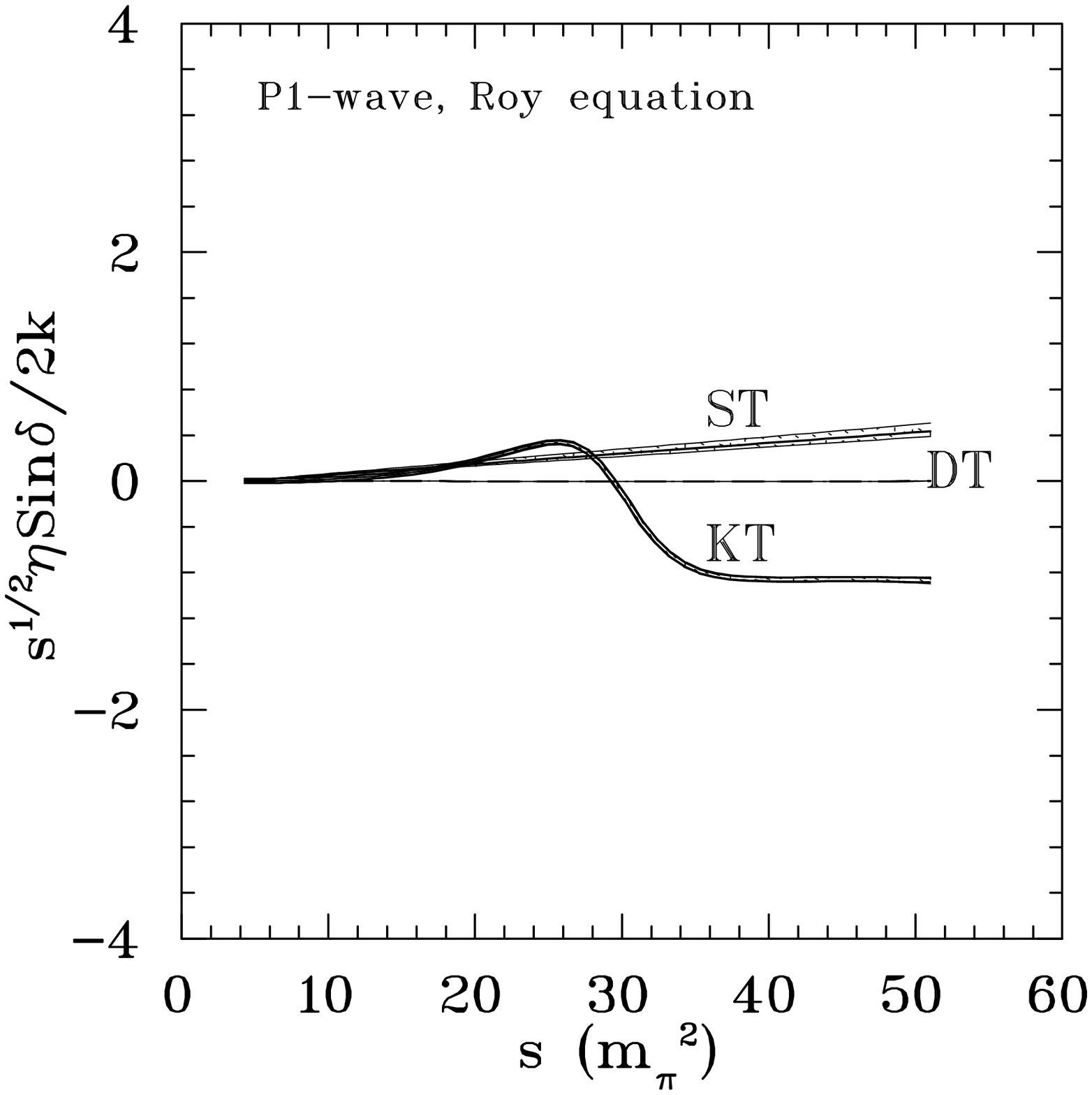}
\includegraphics*[width=6.0cm]{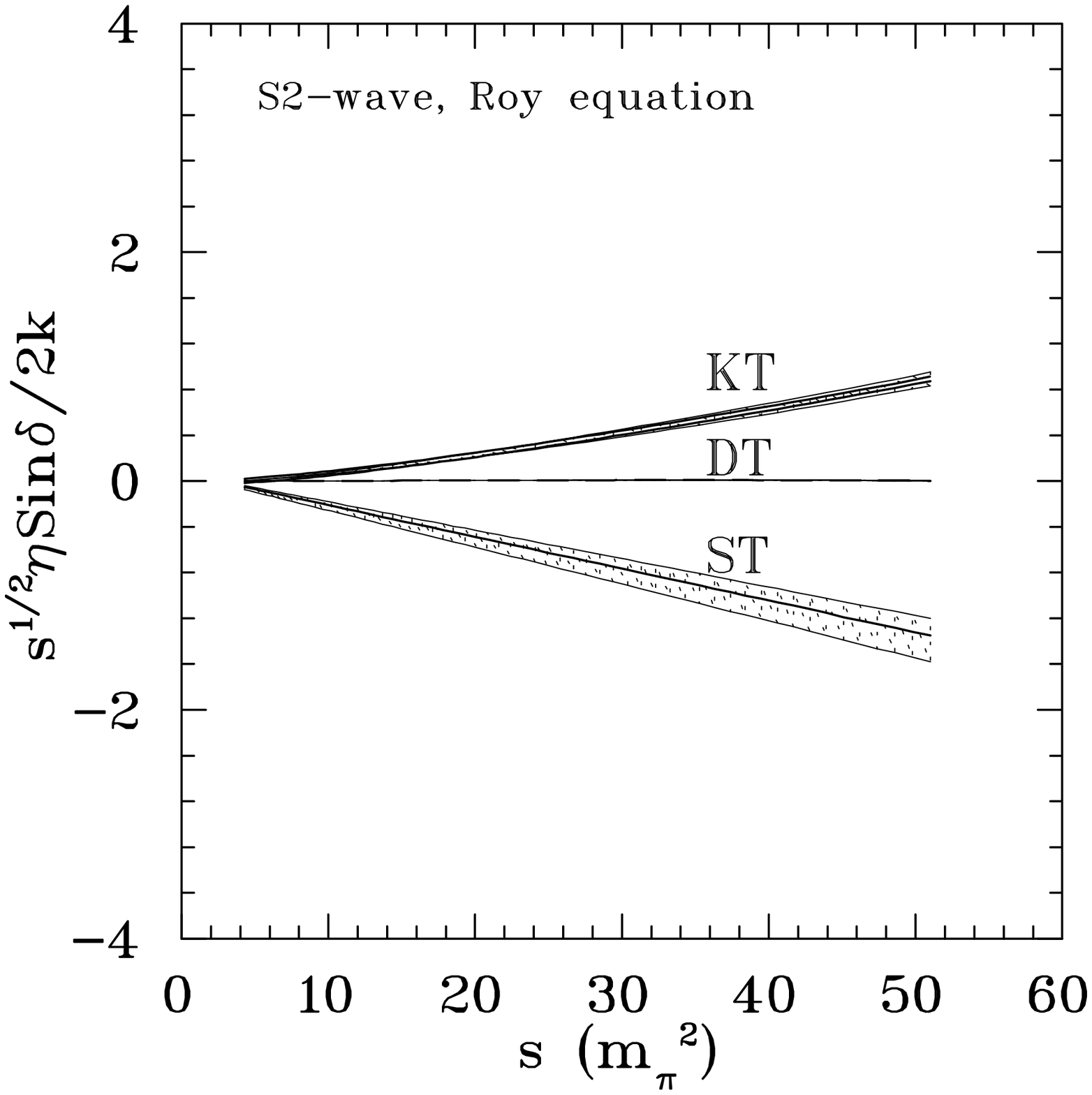}
\vspace{-1cm}
\caption{Subtraction ($ST$), kernel ($KT$) and driving ($DT$) terms for  
the $S0$, $P$ and $S2$ waves from the twice-subtracted Roy's equations.
Dashed bands denote the errors of these terms.}
\label{fig:decompositionRoy}
\end{figure}

\begin{figure}[ht!]
\vspace{9pt}
\vspace{-0.8cm}
\includegraphics*[width=5.9cm]{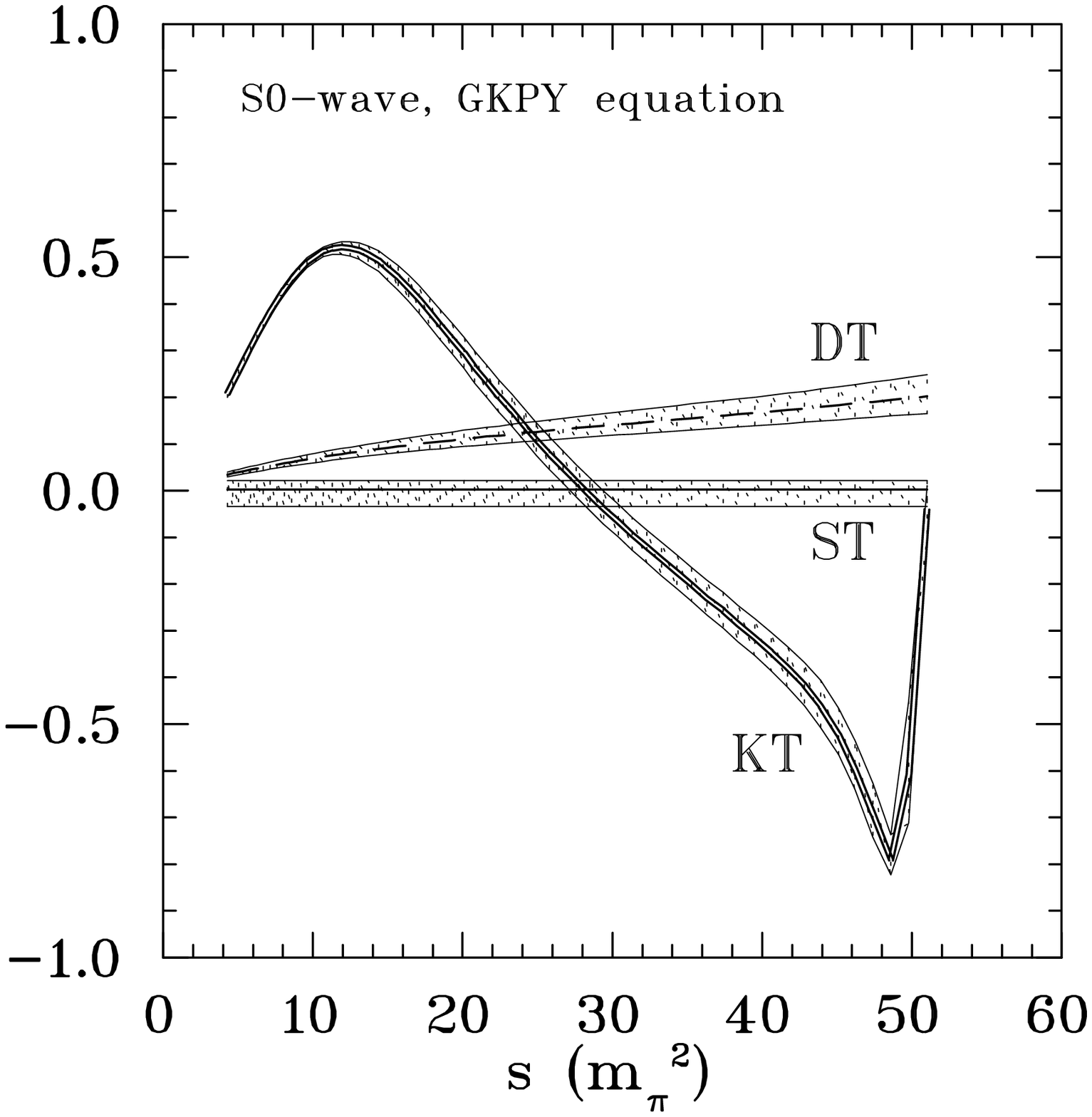}
\includegraphics*[width=5.9cm]{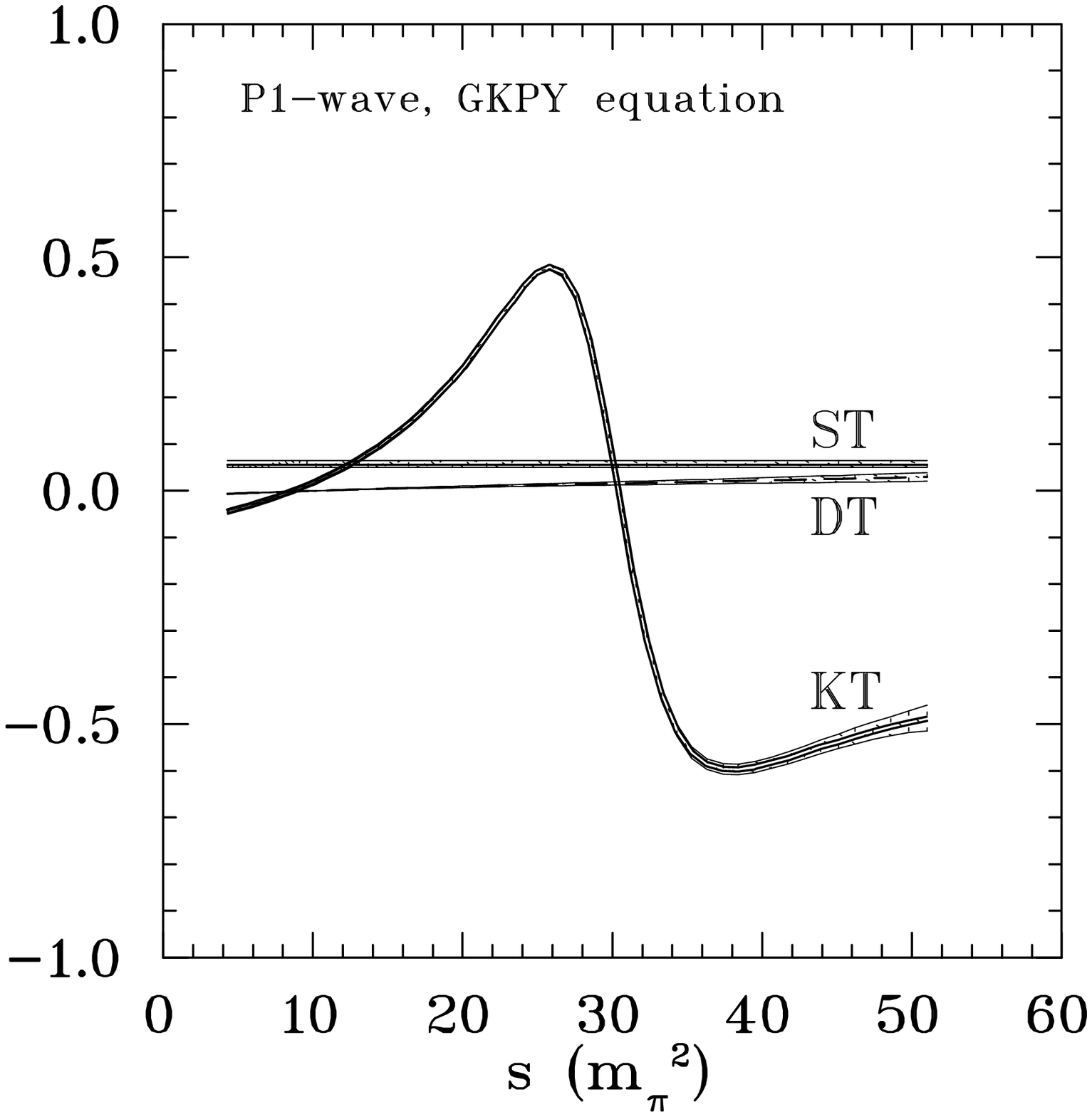}
\includegraphics*[width=5.9cm]{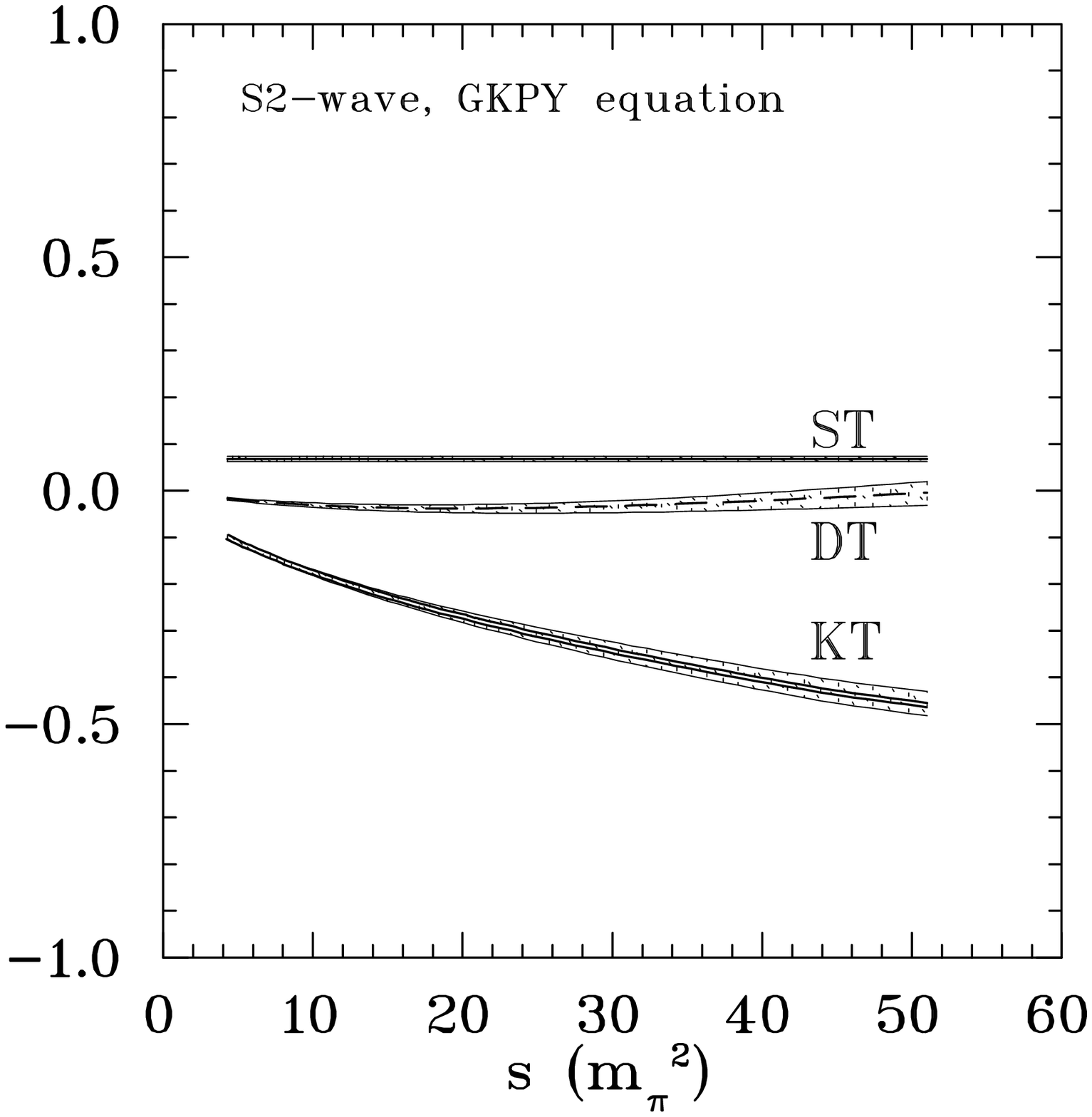}
\vspace{-1cm}
\caption{As in Fig.~\ref{fig:decompositionRoy} but for GKPY equations.}
\label{fig:decompositionPaco}
\end{figure}

By comparing Fig.~\ref{fig:decompositionRoy} for RE with 
Fig.~\ref{fig:decompositionPaco} for GKPY eqs. 
very remarkable differences (especially for the $S0$ wave) 
can be observed in the relative
sizes of the $ST$ and $KT$ terms.
Please note the different scales.
The curves represent the 
real part (output) of the amplitudes, i.e., $s^{1/2}\,\eta\,\sin\,\delta/2k$, where
$\delta$ and $\eta$ are the phase shifts and elasticities for the
preliminary 
amplitudes of the Constrained Data Fit
described in the talk by J.R. Pel\'aez in
this conference.  
This fit describes data and has been constrained to 
satisfy FDR, RE. GKPY eqs. and some crossing sum rules.

Note that, for RE, the $ST$ and $KT$ terms, which are huge,
suffer a strong cancellation. Actually, 
for sufficiently large energy, both the $ST$ and $KT$ terms
are much larger than the unitarity bound $|Re\, t|\leq s^{1/2}/2k\sim1$, which is
only satisfied by the real part of the total amplitude after their
strong cancellation. In contrast, GKPY eqs. are dominated by $KT$ terms,
which are always smaller than one. Remarkably, 
the $DT$ are still relatively small compared with the $KT$ despite
there is one less subtraction in GKPY eqs. This means that the 
effect of high energy input, whose knowledge is less detailed, is well under
control.

The error bands depicted in
Figs.~\ref{fig:decompositionRoy} and ~\ref{fig:decompositionPaco}
are generated using a Monte Carlo Gaussian sampling of
all parameters of the Constrained Fit to Data,
(varied within 3 standard deviations). 
Uncertainties are just the 
widths of independent Gaussian fits
to the left and right sides of the $ST$, $KT$ and $DT$ distributions.
As seen in Fig.~\ref{fig:decompositionRoy}, for sufficiently high energies,
the linear $s$ dependence of the $ST$ terms makes their uncertainty
large and dominant when using RE for the $S0$ and $S2$ waves.
This does not happen for GKPY eqs. whose errors, given the same input, are 
 smaller, above $s \sim 400$ MeV,
than those from RE (see also Fig. 1 in J.R. Pelaez's talk in this conference).
For this reason we are now implementing the 
GKPY eqs. {\it together} with FDR and standard Roy. eqs. 
not only to constraint
the amplitude above $\sim 400$ MeV, but also to extend analytically 
the amplitudes to determine the $\sigma$ pole position.

\section{Position of the $\sigma$ pole}
The mass and width of the $\sigma$ or $f_0(600)$
meson quoted in the
Particle Data Table are very widely spread~\cite{PDG06}
\begin{equation}
M_{\sigma} - i\frac{\Gamma_{\sigma}}{2} \approx (400 - 1200) - i(250 - 500)
\mbox{\boldmath{MeV}}.
\end{equation}
The main reason of these uncertainties is that $\pi\pi$ scattering data
 are few and sometimes contradictory.
Moreover, all the quoted theoretical models are 
not equally reliable, and even less so when extending the amplitude
to the complex plane.
Thus the position of the sigma 
pole in various models differ significantly~\cite{PDG06}, although,
with a couple of exceptions, they tend to agree around
$M_{\sigma} - i\Gamma_{\sigma}/2 \approx (400 - 500) - i(220 - 300)
\mbox{ MeV}$.

The recent data from E865 collaboration at Brookhaven~\cite{Pislak2001} and from NA48/2~\cite{NA482008} 
provide us with new and very precise information on the $\pi\pi$ scattering at low energies.
Thanks to these new data we are 
able to construct, with our Constrained Fits to Data,
a very reliable description
for the $S0$ wave especially near the $\pi\pi$ threshold (see J.R. Pel\'aez
talk in this conference).

With those precise data parametrizations,
we can now use either the Roy eqs. or the GKPY eqs. to
extend the partial waves analytically to the complex plane 
and look for poles in the second sheet of the S-matrix. 
As it well known, a pole on the second 
Riemann sheet (unphysical sheet) is associated with a
zero on the first---the physical one.
Therefore, as usual, we look 
for zeroes of the physical sheet of the $S$-matrix,
\begin{equation}
S_0^0(s) = 1 + 2\,i\,\sqrt{1-4m_{\pi}^2/s}\,f_0^0(s).
\end{equation}
Depending on whether we use Roy or GKPY eqs. we find a different accuracy in our results, namely
\begin{eqnarray}
\hspace*{-.3cm}&\sqrt{s_{\sigma}} =  459^{+36}_{-33} - i\,257^{+17}_{-18} \mbox{ \boldmath{MeV}} \hspace*{.5cm} & {\rm (RE)} 
\label{sigmaRE} \\
\hspace*{-.3cm}&\sqrt{s_{\sigma}} =  461^{+14.5}_{-15.5} - i\,255\pm 16 \mbox{ \boldmath{MeV}} 
& {\rm (GKPY)}\hspace*{.5cm}
\label{sigmaGKPY}  
\end{eqnarray}
from the GKPY equations.
These values are in good agreement with each other. 
Note that the errors have been calculated in the same way as for the output
amplitudes in Sect. 1, i.e. from a Monte Carlo Gaussian sampling.
Except for those of the mass from RE, which are roughly 25\% smaller,
they are almost identical to the non-Gaussian symmetric errors that we 
have been using in previous works \cite{RUBENCRACOVIA}
\begin{equation}
\sqrt{s_{\sigma}} =(461\pm14) - i\,(255\pm 16)\, {\rm MeV}\quad{\rm (GKPY)}
\end{equation}
It is reassuring that our error estimates remain stable calculating them
in these two rather different ways.

This preliminary determination
is quite consistent with other recent and precise results in the literature: 
On the one hand, both the mass and width
lie less than 1.25 standard deviations 
from the prediction of 
twice-subtracted Roy's equations combined with ChPT results
for the scattering lengths \cite{Caprini}: 
$\sqrt{s_{\sigma}} = 441^{+16}_{-8} -i\,272^{+9}_{-14.5} 
\mbox{\boldmath{MeV}}$.
On the other hand our preliminary mass determination above
is also slightly beyond one standard deviation 
from the pole in our simple fit \cite{Yndurain:2007qm}
of a conformal expansion to low energy data 
$\sqrt{s_{\sigma}} = (484\pm17) -i\,(255\pm10) 
\mbox{\boldmath{MeV}}$, but agrees remarkably well on the width.

\section{Conclusions}
The GKPY equations -- Roy-like dispersion relations with one subtraction 
for the $\pi\pi$ amplitudes -- provide stringent
constraints for dispersive analysis of experimental data.
The main advantage of GKPY eqs. is that, for the same input,
 in the $0.4 \mbox{ GeV}\,\leq\sqrt{s}\leq 1.1\mbox{ GeV}$
region they have  significantly smaller errors than standard Roy. eqs.
Hence, they provide better accuracy tests 
and analytic extensions of the amplitudes in that region.
In particular, using just a data analysis consistent within errors 
with Forward Dispersion Relations, Roy eqs. and GKPY eqs. (and no ChPT input),
we have presented here the following preliminary but very precise determination
of the $\sigma$ pole position: 
\begin{equation}
\sqrt{s_{\sigma}} =  461^{+14.5}_{-15.5} - i\,255\pm 16 \mbox{ \boldmath{MeV}}
\end{equation}
The details of GKPY eqs. together with a 
full constrained dispersive analysis the data,
extending both GKPY and Roy eqs. up to 1.1 GeV and
some slight improvements on the partial waves
is about to be finished. 
A final number for the $\sigma$ pole should follow relatively
easy once that data analysis is completed.

\section{Acknowledgments}
This talk is dedicated to the memory of F.J. Yndurain.
Partial financial support from Spanish contracts PR27/05-13955-BSCH, FPA2004-02602, UCM-CAM
910309 and BFM2003-00856.


\begin{thebibliography}{9}

\bibitem{Roy71} S.M. Roy, Phys. Lett. B 36, 353 (1971).
\bibitem{A4} B. Ananthanarayan {\it et al.}, Phys. Rept. 353, 207 (2001). 

\bibitem{CGLNPB01} G. Colangelo, J. Gasser and H. Leutwyler, Nucl. Phys. B 603, 125 (2001).

\bibitem{DescotesGenon:2001tn}
  S.~Descotes-Genon, N.~H.~Fuchs, L.~Girlanda and J.~Stern,
  Eur.\ Phys.\ J.\  C {\bf 24}, 469 (2002)

\bibitem{KJYII} R. Kami\'nski, J.R. Pelaez and F.J. Yndurain, Phys. Rev. D 77, 054015 (2008). 
\bibitem{KJYI} R. Kaminski, J.R. Pelaez and F.J. Yndurain, Phys. Rev. D 74, 014001 (2006).

 
  
\bibitem{Kaminski:2002pe}
  R.~Kaminski, L.~Lesniak and B.~Loiseau,
  Phys.\ Lett.\  B {\bf 551}, 241 (2003)



\bibitem{Caprini} I. Caprini, G. Colangelo and H. Leutwyler, Phys. Rev. Lett. 96, 132001 (2006).

\bibitem{PDG06} W. M. Yao {\it et al.}, J. Phys. G33, 1-1232 (2006).
\bibitem{Pislak2001} S. Pislak {\it et al.}, Phys. Rev. Lett. 87, 221801 (2001).
\bibitem{NA482008} J. R. Batley {\it et al.}, CERN-PH-EP-2007-035 (2008).


\bibitem{RUBENCRACOVIA}
R.~Garcia-Martin {\em et al.}, To appear in the proceedings of the
Meson08 conference, Cracow, Poland.


\bibitem{Yndurain:2007qm}
  F.~J.~Yndurain, R.~Garcia-Martin and J.~R.~Pelaez,
  Phys.\ Rev.\  D {\bf 76}, 074034 (2007)
\end{thebibliography}
\end{document}